\documentclass{svmult}
\usepackage{amssymb,amsmath} 
\usepackage{graphicx}

\begin{document}

\def\={\equiv} 
\def\div{\nabla\cdot } 
\def\pl{\partial}
\def\since{\ \because\ }
\def\tf{\ \therefore\ }

\newcommand{\cast}{\circledast}
\newtheorem{thm}{Theorem}
\newcommand{\bib}{\bibitem}
\newcommand{\mrg}[1]{{\mathring#1}}
\newcommand{\tg}[1]{{\tag{$#1$}}}
\newcommand{\nt}{\notag}
\newcommand{\ci}{\cite}
\newcommand{\lab}{\label}
\newcommand{\tlab}{\label}
\newcommand{\eq}{\eqref}

\newcommand{\bl}{\renewcommand{\baselinestretch}} 
\newcommand{\cl}{\centerline}

\newcommand{\bx}[1]{ \boxed{#1}}
\newcommand{\bbx}[1]{\boxed{\boxed{\ #1\ }}}

\newcommand{\lp}{\left(}
\newcommand{\rp}{ \right)}
\newcommand{\lb}{ \left[}
\newcommand{\rb}{\right]}
\newcommand{\la}{\langle\,}
\newcommand{\LA}{\left\langle\,}
\newcommand{\ra}{\,\rangle}
\newcommand{\RA}{\right\rangle\,}
\newcommand{\LB}{\left\lbrace}
\newcommand{\RB}{\right\rbrace}

\newcommand{\ola}[1]{\overleftarrow{#1}} 
\newcommand{\olra}[1]{\overleftrightarrow{#1}} 

\newcommand{\0}[1]{{(#1)}}
\newcommand{\1}[1]{{\hat #1}}
\newcommand{\2}[1]{{\tilde #1}}
\newcommand{\3}[1]{{\boldsymbol #1}}
\newcommand{\bb}[1]{{\boldsymbol{\bar #1}}}
\newcommand{\bh}[1]{{\boldsymbol{\hat #1}}}
\newcommand{\bt}[1]{{\boldsymbol{\tilde #1}}}
\newcommand{\bfs}[1]{{\mathbf#1}}
\newcommand{\btil}[1]{{\mathbf{\tilde{ #1}}}}
\newcommand{\bhat}[1]{{\mathbf{\hat{ #1}}}}
\newcommand{\4}[1]{{\,\mathbb#1}}
\newcommand{\bbtil}[1]{{\,\mathbb{\tilde{ #1}}}}
\newcommand{\5}[1]{{\mathcal#1}}
\newcommand{\ctil}[1]{{\mathcal{\tilde{ #1}}}}
\newcommand{\chat}[1]{{\mathcal{\hat{ #1}}}}
\newcommand{\fr}[1]{{\mathfrak #1}}
\newcommand{\6}[1]{_{\scriptscriptstyle#1}}
\newcommand{\7}[1]{{\bar#1}}
\newcommand{\8}{\infty}
\newcommand{\9}[1]{^{\,\scriptscriptstyle#1}}
\newcommand{\cd}[1]{{\lfloor #1\rfloor}}

\newcommand{\bbh}[1]{{\mathbb{\widehat#1}}}
\newcommand{\bbt}[1]{{\mathbb{\widetilde#1}}}
\newcommand{\w}{\hskip.5pt\9\flat}
\newcommand{\wh}[1]{{{\widehat#1}}}
\newcommand{\wt}[1]{\widetilde{#1}}

\newcommand{\dtb}[1]{\dot{{\boldsymbol #1}}}
\newcommand{\ddtb}[1]{\ddot{{\boldsymbol #1}}}


\def\a{\alpha} 
\def\b{\beta} 
\def\be{\backepsilon}
\def\c{\chi}
\def\d{\delta} 
\def\e{\varepsilon} 
\def\ee{\epsilon}
\def\f{\phi} 
\def\vf{\varphi} 
\def\g{\gamma}
\def\h{\eta} 
\def\ii{\iota}
\def\k{\kappa}
\def\vk{\varkappa}
\def\l{\lambda} 
\def\m{\mu} 
\def\n{\nu}
\def\o{\omega} 
\def\p{\pi} 
\def\vp{\varpi}
\def\q{\theta} 
\def\vq{\vartheta} 
\def\r{\rho}
\def\vr{\varrho} 
\def\s{{\sigma}} 
\def\vs{\varsigma}
\def\t{\tau} 
\def\u{\upsilon} 
\def\x{\xi}
\def\y{\psi} 
\def\z{\zeta}

\def\D{\Delta} 
\def\F{\Phi} 
\def\G{\Gamma}
\def\L{\Lambda} 
\def\O{\Omega} 
\def\P{\Pi}
\def\Q{\Theta} 
\def\S{\Sigma} 
\def\U{\Upsilon}
\def\Y{\Psi}
\def\X{\Xi} 


\newcommand{\hb}[1]{{\ \text{#1}\ }}
\newcommand{\bul}{$\bullet\ $}
\newcommand{\cir}{$\circ\ $}
\newcommand{\db}{{\,{\rm d}\kern-.9ex {^-}}\!}
\newcommand{\dir}{{\pl\kern-1.2ex {/}}}
\newcommand{\dd}{{\,\rm d}}
\newcommand{\Star}{$\bigstar$}
\newcommand{\bs}{\backslash}

\newcommand{\app}{\approx} 
\newcommand{\cc}[1]{{{\mathbb C\hskip.5pt}^{#1}}}
\newcommand{\ccc}[1]{{\mathbb C\hskip.5pt}^{#1+1}}
\newcommand{\const}{{\rm constant}}
\newcommand{\curl}{\nabla\times} 

\newcommand{\eg}{{\it e.g., }}
\newcommand{\grad}{\nabla} 
\newcommand{\hra}{\hookrightarrow}
\newcommand{\ie}{{\it i.e., }}
\newcommand{\Iff}{\ \Longleftrightarrow\ }
\newcommand{\llra}{\ \longleftrightarrow\ }
\def\iff{\ \Leftrightarrow\ }
\newcommand{\im}{{\,\rm Im}\ }  
\newcommand{\imp}{\ \Rightarrow\ }
\newcommand{\inv}{^{-1}}
\newcommand{\intt}{\int\!\!\!\int}
\newcommand{\ir}{\int_{-\infty}^\infty} 
\newcommand{\llr}{L^2(\3R)}
\newcommand{\lra}{\leftrightarrow}
\newcommand{\Lra}{\Leftrightarrow} 
\newcommand{\Llra}{\Longleftrightarrow}
\newcommand{\Log}{{\rm Log\,}}
\newcommand{\mink}{\4R^{3,1}}
\newcommand{\minkd}{\4R_{3,1}}
\newcommand{\mt}{{\,\mapsto\,}}
\newcommand{\nm}{{n-1}}
\newcommand{\nn}{{n+1}}
\newcommand{\ptl}[2]{{\frac{\partial #1}{\partial#2}}}
\newcommand{\plra}{\pl^{\kern-1.25ex^\lra}}
\newcommand{\qq}{\quad} 
\newcommand{\qqq}{\qquad} 
\newcommand{\re}{{\,\rm Re}\  }   
\newcommand{\rr}[1]{{{\mathbb R}^{#1}}}
\newcommand{\rrr}[1]{{\mathbb R}^{#1+1}}
\newcommand{\sea}{\searrow}
\newcommand{\sgn}{{\,\rm Sgn \,}}
\newcommand{\sh}[1]{\hskip#1ex} 
\newcommand{\sr}{\sqrt}
\newcommand{\stp}{\bf STOP \rm}
\newcommand{\supp}{{\rm supp \,}}
\newcommand{\sv}[1]{\vskip#1ex}
\def\tr{{\rm tr\,}}

\newcommand{\tpi}{(2\p i)}
\newcommand{\vv}{{\rm v}}
\newcommand{\ww}{{\rm w}}
\newcommand{\oo}{\cdot\3\o}
\newcommand{\st}{\cdot\3s dt}
\newcommand{\xt}{\cdot d\3x dt}
\newcommand{\xx}{\cdot d\3x}

\newcommand{\orr}{{(\3r)}}
\newcommand{\ort}{{(\3r,t)}}
\newcommand{\ozt}{{(\3z,\t)}}

\def\Xint#1{\mathchoice
   {\XXint\displaystyle\textstyle{#1}}%
   {\XXint\textstyle\scriptstyle{#1}}%
   {\XXint\scriptstyle\scriptscriptstyle{#1}}%
   {\XXint\scriptscriptstyle\scriptscriptstyle{#1}}%
   \!\int}
\def\XXint#1#2#3{{\setbox0=\hbox{$#1{#2#3}{\int}$}
     \vcenter{\hbox{$#2#3$}}\kern-.5\wd0}}
\def\ddashint{\Xint=}
\def\ppint{\Xint-}

\def\HB{\hfill\break}
\def\VE{\vfill\eject}

\def\bib#1{\bibitem[#1]{#1}}

\title*{Eigenwavelets of the Wave Equation}
\toctitle{Eigenwavelets of the Wave Equation}
\titlerunning{Eigenwavelets of the Wave Equation}
\author{Gerald Kaiser}
\authorrunning{G. Kaiser}

\institute{Signals \& Waves, Austin, TX $\bullet$ www.wavelets.com  $\bullet$
\texttt{kaiser@wavelets.com}}

\maketitle

\begin{abstract}
\noindent 
We study a class of localized solutions of the wave equation, called \sl eigenwavelets, \rm obtained by extending its fundamental solutions to complex spacetime in the sense of hyperfunctions. The imaginary spacetime variables $y$, which form a timelike vector, act as \sl  scale parameters \rm generalizing the scale variable of wavelets in one dimension. They determine the \sl shape \rm of the wavelets in spacetime, making them \sl pulsed beams \rm that can be focused as tightly as desired around a single ray by letting $y$ approach the light cone. Furthermore, the absence of any sidelobes makes them especially attractive for communications, remote sensing and other applications using acoustic waves. (A similar set of `electromagnetic eigenwavelets' exists for Maxwell's equations.) I review the basic ideas in Minkowski space  $\4R^{3,1}$, then compute sources whose realization should make it possible to radiate and absorb such wavelets. This motivates an extension of Huygens' principle allowing equivalent sources to be represented on shells instead of surfaces surrounding a bounded source. 

\end{abstract}

\section{Extension of wave functions to complex spacetime}

The ideas to be presented here affirm that complex analysis resonates deeply in ``real'' physical and geometric settings, and so they are close in spirit to the work of Carlos Berenstein (see \ci{BG91,BG95,B98} for example), to whom this volume is dedicated. 

Acoustic and electromagnetic wavelets were first constructed in \ci{K94}. It was shown that solutions of \sl homogeneous \rm (\ie sourceless) scalar and vector wave equations in Minkowski space $\4R^{3,1}$ extend naturally to complex spacetime, and the wavelets were defined as the Riesz duals of \sl evaluation maps \rm acting on spaces of such holomorphic solutions. The sourceless wavelets then split naturally into retarded and advanced parts emitted and absorbed, respectively, by sources located on \sl branch cuts \rm needed to make these parts single-valued. Later work \ci{K3, K4} was aimed at the construction of \sl realizable \rm source distributions which, when synthesized, would act as antennas radiating and receiving the wavelets. Two difficulties with this approach have been (a) that the computed sources are quite singular, consisting of multiple surface layers that may be difficult to realize in practice, and (b) in the electromagnetic case the sources appeared to require a nonvanishing magnetic charge distribution, which cannot be realized as no magnetic monopoles have been observed in Nature. In this paper we resolve the first difficulty by replacing the spheroidal surface supporting the sources in \ci{K3, K4} by a spheroidal \sl shell. \rm  It is shown in \ci{K4a} that the second difficulty can be overcome using Hertz potentials, which give a charge-current distribution due solely to \sl bound electric charges \rm confined to the shell.

Although our constructions generalize to other dimensions, we shall concentrate here on the physical case of the Minkowski space $\4R^{3,1}$. Let
\begin{align}\lab{xy}
x&=(\3r,t), \  y=(\3a,b)\in\4R^{3,1}
\end{align}
be real spacetime vectors and define the complex \sl causal tube \rm
\begin{align}\lab{T}
\5T=\{x-iy\in\cc4:\ y\hb{is timelike, \ie} |b|>|\3a|\}.
\end{align}
It was shown in \ci{K94, K3} that solutions of the \sl homogeneous \rm wave equation
\begin{align}\lab{hom}
\Box f_0\0x \=(\pl_t^2-\D)f_0(\3r,t)=0
\end{align} 
extend naturally to analytic functions $\2f_0(x-iy)$ in $\5T$ in the sense that
\begin{align}\lab{lim}
\lim_{y\to+0}\LB \2f_0(x-iy)-\2f_0(x+iy)\RB=f_0\0x,
\end{align}
where $y\to +0$ means that $y$ approaches the origin within the future cone, \ie with $b>|\3a|$. This kind of extension to complex domains is familiar in hyperfunction theory; see \ci{K88,KS99} for example.  We now show that even when the wave function has a source,  \ie
\begin{align}\lab{inhom}
\Box f\0x=4\p g\0x,
\end{align}
it extends analytically to $\5T$ outside a spacetime region determined by the source.  It will suffice to do this for the \sl retarded propagator \rm 
\begin{align}\lab{prop}
G\0x=\frac{\d(t-r)}r,
\end{align}
which is the unique causal fundamental solution: 
\begin{align}\lab{Y0}
\Box G\0x=4\p\d\0t\d(\3r)=4\p\d\0x,\qq G(\3r,t)=0\ \forall t<0.
\end{align}
If the source $g$ is supported in a compact spacetime region $W$, the unique causal solution of  \eq{inhom} is given by
\begin{align}\lab{f1}
f\0x=\int_W dx'\ G(x-x') g(x').
\end{align}
Assume for the moment that $G\0x$ has been extended to $\2G(x-iy)$. Then we \sl define \rm  the source of $\2G$ as the distribution $\2\d$ in \sl real \rm spacetime given by
\begin{align}\lab{2d}
4\p\2\d(x-iy)\=\Box_x\2G(x-iy),
\end{align}
where $\Box_x$ means that the wave operator acts only on $x$, in a distributional sense, so that the imaginary spacetime vector $y$ is regarded as an auxiliary parameter. The extended solution is now defined as
\begin{align}\lab{f2}
\2f(x-iy)=\int_W dx'\ \2G(x-x'-iy) g(x')
\end{align}
and it satisfies the wave equation
\begin{align*}
\Box_x\2f(x-iy)=4\p\2g(x-iy)
\end{align*}
with the extended source
\begin{align}\lab{g2}
\2g(x-iy)=\int_W dx'\ \2\d(x-x'-iy) g(x').
\end{align}
\sl Formally, \rm the extended delta function $\2\d(x-iy)$ is a `point source' at the imaginary spacetime point $iy$ as seen by a real observer at $x$. Actually, it will be seen to be a distribution in $x$ with compact \sl spatial \rm (but not temporal) support localized around the spatial origin $(\3r=\30$) and depending on the choice of a \sl branch cut \rm needed to make $\2G$ single-valued. This branch cut is precisely the region where $\2G$ fails to be analytic, and the integral \eq{f2} determines a region $\2W$ containing $W$ where $\2f$ fails to be analytic. 

A \sl general \rm solution $f_1\0x$ of \eq{inhom} is obtained by adding a sourceless wave $f_0\0x$ to \eq{f1}. Since $\2f_0$ is analytic in $\5T$, $\2f_1$ is analytic in $\5T$ outside of $\2W$.  It therefore suffices to concentrate on the propagators as claimed. In the rest of the paper we construct extended propagators, study their properties, and compute their sources.

\section{Extended propagators}

In accordance with \eq{xy}, we use the following notation for complex space and time variables:
\begin{align*}
\bt r&=\3r-i\3a\in\cc3,\qqq \2t=t-ib\in\4C\\
\2x&=x-iy=(\bt r, \2t)\in\5T\iff |b|>|\3a|.
\end{align*}
As above, we interpret $i\3a$ formally as an imaginary spatial source point, so that $\bt r$ is the vector from the imaginary source point $i\3a$ to a real observer at $\3r$. To extend the propagator \eq{prop}, begin by replacing the one-dimensional delta function with the \sl Cauchy kernel, \rm
\begin{align}\lab{cauchy}
\d\0t\to\2\d(\2t)=\frac1{2\p i\2t}\,, \qq \2t=t-ib,
\end{align}  
which indeed satisfies a condition of type \eq{lim}:
\begin{align}\lab{lim2}
\lim_{b\to +0}\LB\2\d(t-ib)-\2\d(t+ib)\RB=\d\0t.
\end{align}
To complete the extension of $G(\3r,t)$, we must also extend the Euclidean distance $r(\3r)=|\3r|$.  Define the \sl complex distance \rm from the source to the observer as
\begin{align}\lab{sig}
\2r(\bt r)=\sr{\bt r\cdot\bt r}=\sr{r^2-a^2-2i\3r\cdot\3a}, \ \hb{where}\ r=|\3r|,\ a=|\3a|.\,
\end{align}
$\2r(\bt r)$ is an analytic continuation to $\cc3$ of $r(\3r)$. 
Being a complex square root, it has branch points wherever 
$\bt r\cdot\bt r=0$. For fixed $\3a\ne \30$,  these form a circle of radius $a$ in the plane orthogonal to $\3a$,\footnote{In $\rr n$, $\5C$ would be a sphere of codimension 2 orthogonal to $\3a$.}
\begin{align}\lab{C}
\5C\=\{\3r\in\rr3: \2r=0\}=\{\3r:\  r=a,\  \3r\cdot\3a=0\}.
\end{align}
To be consistent with the notation $\bt r=\3r-i\3a$, we write
\begin{align}\lab{pq}
\2r=p-iq.
\end{align}
Comparison with \eq{sig} gives the following relations between $(p,q)$ and the spherical and cylindrical coordinates with axis along $\3a$:
\begin{align}\lab{pq1}
p^2-q^2=r^2-a^2,\qqq pq=\3a\cdot\3r=ar\cos\q=az
\end{align}
and 
\begin{align}
a^2\r^2&=a^2(r^2-z^2)=a^2(a^2+p^2-q^2)-p^2q^2\nt\\
&=(a^2+p^2)(a^2-q^2). \lab{pq2}
\end{align}
It follows that the real and imaginary parts of $\2r$ are bounded by $r$ and $a$, respectively:
\begin{align}\lab{qa}
&  p^2\le r^2, \ \hb{\ie } \  | \re\2r |\le |\re\bt r|  \nt  \\
&  q^2\le a^2, \ \hb{\ie } \  | \im\2r |\le |\im\bt r|,
\end{align}
with equalities attained only when $\3r$ is parallel or antiparallel to $\3a$.

Since $\3a$ will be a fixed nonzero vector throughout, we will usually regard $\2r, p, q$  as functions of $\3r$ only, suppressing the dependence on $\3a$. Note that $\rr3-\5C$ is multiply connected since a closed loop that threads $\5C$ cannot be shrunk continuously to a point without intersecting $\5C$. In particular, if we continue $\2r$ analytically around a simple closed loop,  we obtain the value $-\2r$ instead of $\2r$ upon returning to the starting point. Thus $\2r$ is a double-valued function on $\rr3$. To make it single-valued, we choose a branch cut that must be crossed to close the loop. Instead of returning to the starting point as $-\2r$, the sign reversal now takes place upon crossing the cut.  To give an extension of the \sl positive \rm distance, the branch must be chosen so that
\begin{align}\lab{pos}
\3a\to\30\imp \2r\to +r,
\end{align}
and the simplest such choice is obtained by requiring  
\begin{align}\lab{ppos}
\re\2r=p\ge 0.
\end{align}
The resulting branch cut consists of the disk spanning the circle $\5C$, 
\begin{align}\lab{D}
\5D\=\{\3r\in\rr3: p=0\}=\{\3r: \ r\le a,\ \3r\cdot\3a=0\}, \qq \pl\5D=\5C.
\end{align}
$\5D$ will be called the \sl standard branch cut \rm and  $\2r$ the \sl standard complex distance. \rm General branch cuts, obtained by deforming $\5D$ while leaving its boundary intact, will be considered in the next section.

\begin{figure}[ht]
\begin{center}
\includegraphics[width=3 in]{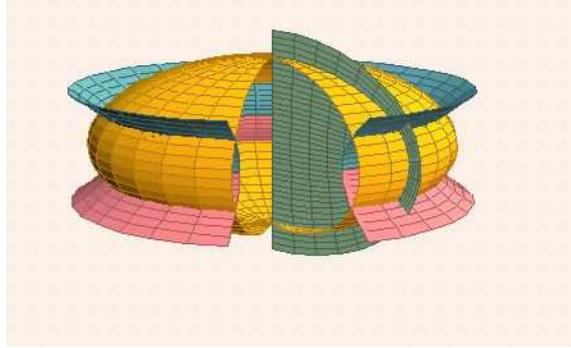}
\caption{The level surfaces of $p, q$ and $\f$ form an oblate spheroidal coordinate system.}
\label{FigOSCS}
\end{center}
\end{figure}

If the observer is far from $\5C$, it follows from \eq{sig} and \eq{pos} that
\begin{align}\lab{far}
r\gg a\imp  p\app r  \hb{\ and\ } q\app a\cos\q,
\end{align} 
Thus, $(p, q/a)$ are deformations of the spherical coordinates $(r, \cos\q)$ near the source. From \eq{pq1} and \eq{pq2} it follows that level surfaces of $p^2$  (as a function of $\3r$, keeping $\3a\ne\30$ fixed)  are spheroids $\5S_p$ and those of $q^2$ are the orthogonal hyperboloids $\5H_q$, given by
\begin{align}
&\5S_p:\qq\frac{\r^2}{p^2+a^2}+\frac{z^2}{p^2}=1,\qq p\ne 0\lab{Sp}\\
&\5H_q:\qq\frac{\r^2}{q^2-a^2}-\frac{z^2}{q^2}=1,\qq 0<q^2<a^2.\lab{Hq}
\end{align}
All these quadrics are \sl confocal \rm  with $\5C$ as the common focal set. As $p\to 0$, $\5S_p$ collapses to a double cover of the disk $\5D$. The variables $(p,q)$, together with the azimuthal angle $\f$ about the $\3a$-axis, determine an \sl oblate spheroidal coordinate system,  \rm as depicted in Figure \ref{FigOSCS}.

We now define the extended propagator as
\begin{align}\lab{2G}
\2G(\bt r, \2t)=\frac{\2\d(\2t-\2r)}{\2r}=\frac1{2\p i\2r(\2t-\2r)}.
\end{align}
This is our \sl basic wavelet,\rm\footnote{In applications, it is better to use time derivatives of $\2G$, which have vanishing moments and better temporal decay and propagation properties \ci{K4}.}
 from which the entire wavelet family is obtained by spacetime translations:
\begin{align}\lab{Yz}
\2G_z\0x=\2G(x-z),\qq z=x'+iy=(\3r'+i\3a, t'+ib)\in\5T
\end{align}
The family $\2G_z$ may be called \sl eigenwavelets \rm of the wave equation in the sense that they are \sl proper \rm to that equation, though of course they are not eigen\sl functions. \rm In fact, $\2G_z\0x$ is seen \ci{K3} to be a \sl pulsed beam \rm originating from $\3r=\3r'$ at $t=t'$ and  propagating along the direction of $\3a/b$, \ie along $\3a$ if $y$ is in the future cone and along $-\3a$ if $y$ is in the past cone. The pulse has a duration $|b|-a$ along the beam axis. By letting $y$ approach the light cone ($a\to |b|$), the beam can be focused as tightly as desired around its axis, approximating a single \sl ray \rm along $y$. Equation \eq{f2} states that the extended causal solution $\2f(x-iy)$ is a superposition of eigenwavelets, all with the same $y$. This gives a \sl directional scale  analysis \rm  of the original solution $f\0x$ which may be called its \sl eigenwavelet transform. \rm

\begin{figure}[ht]
\cl{\includegraphics[width=1.9 in]{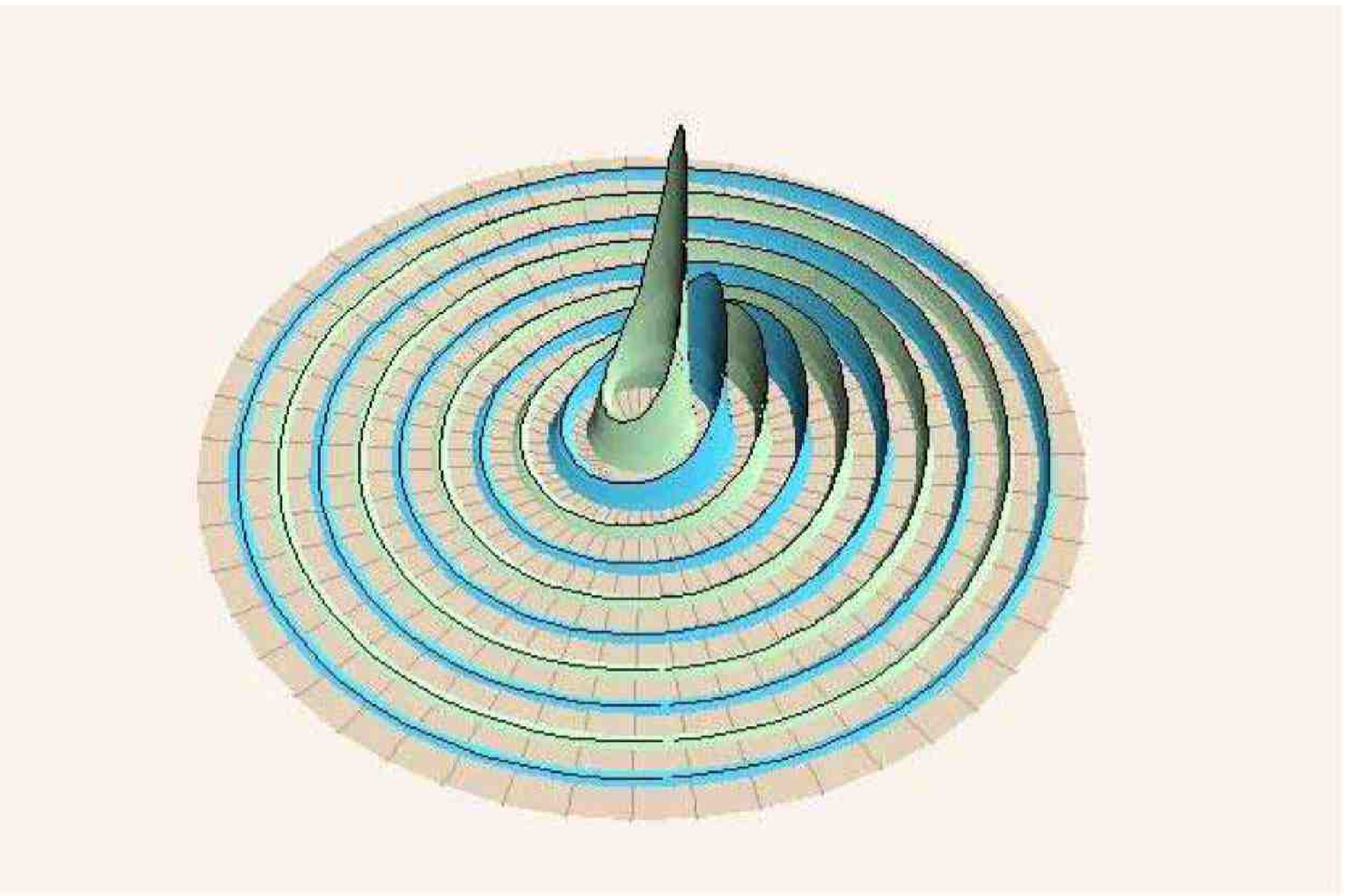}
\includegraphics[width= 1.9  in]{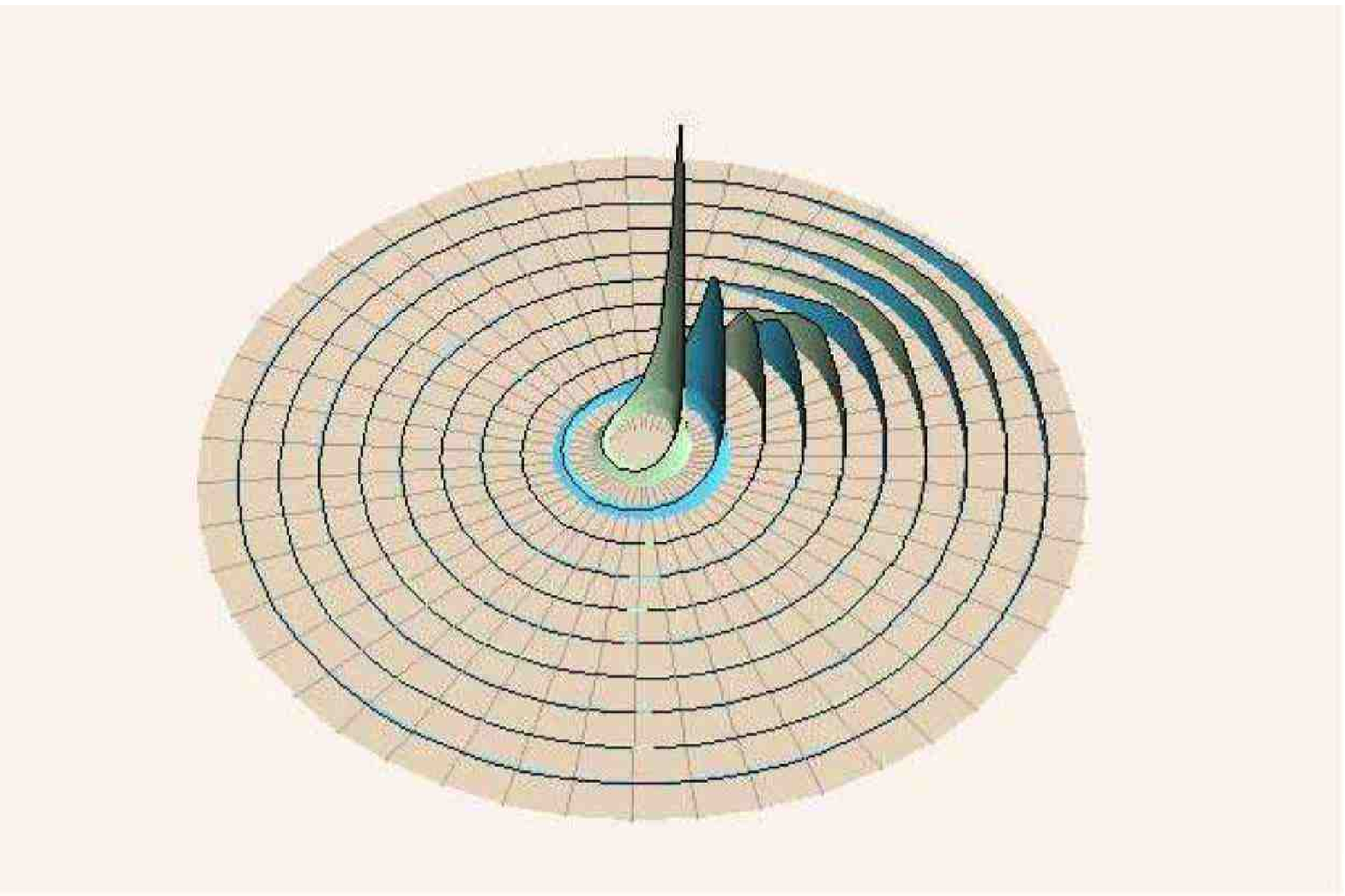}}
\cl{\includegraphics[width= 1.9  in]{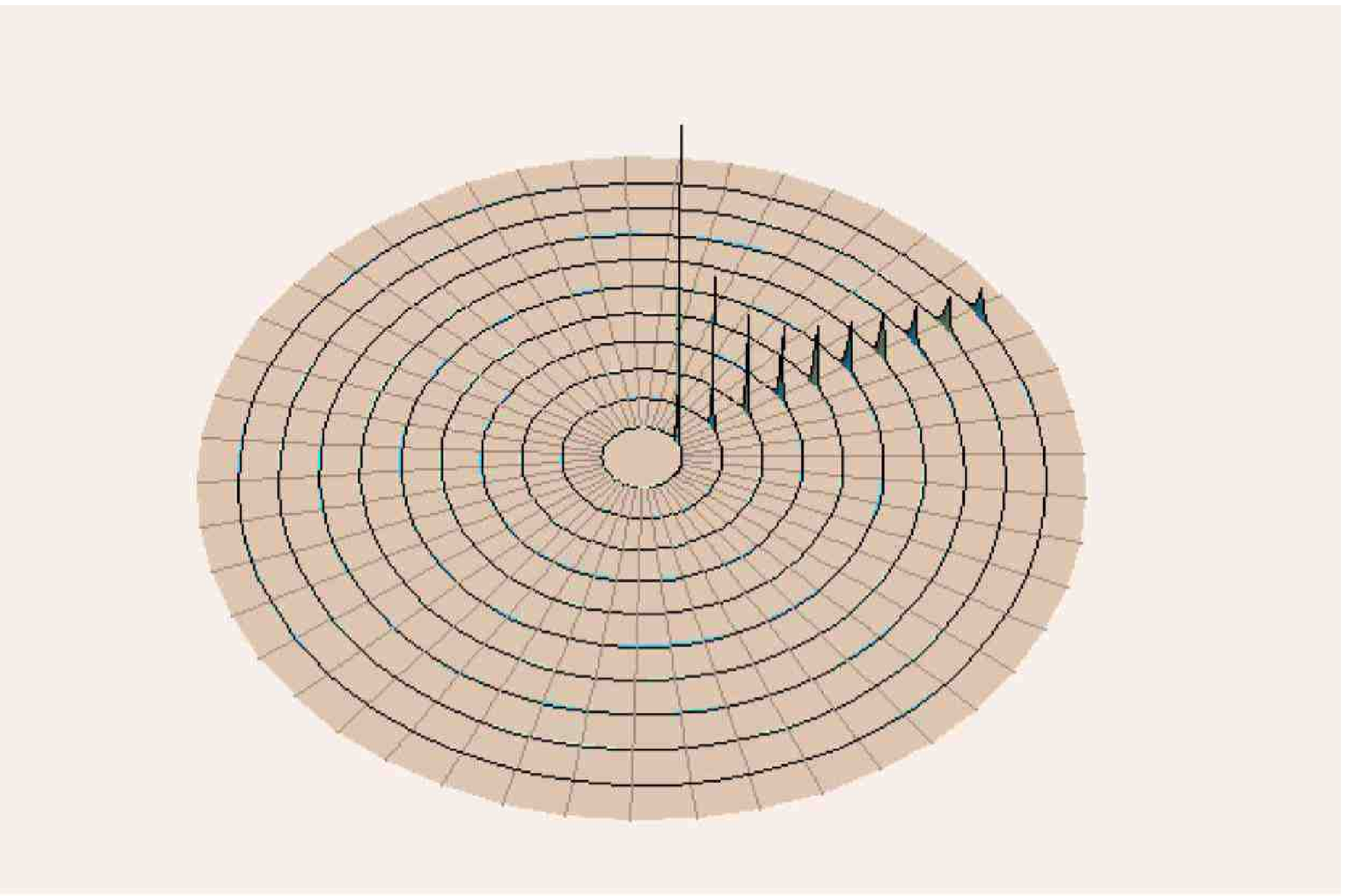}
\includegraphics[width= 1.9 in]{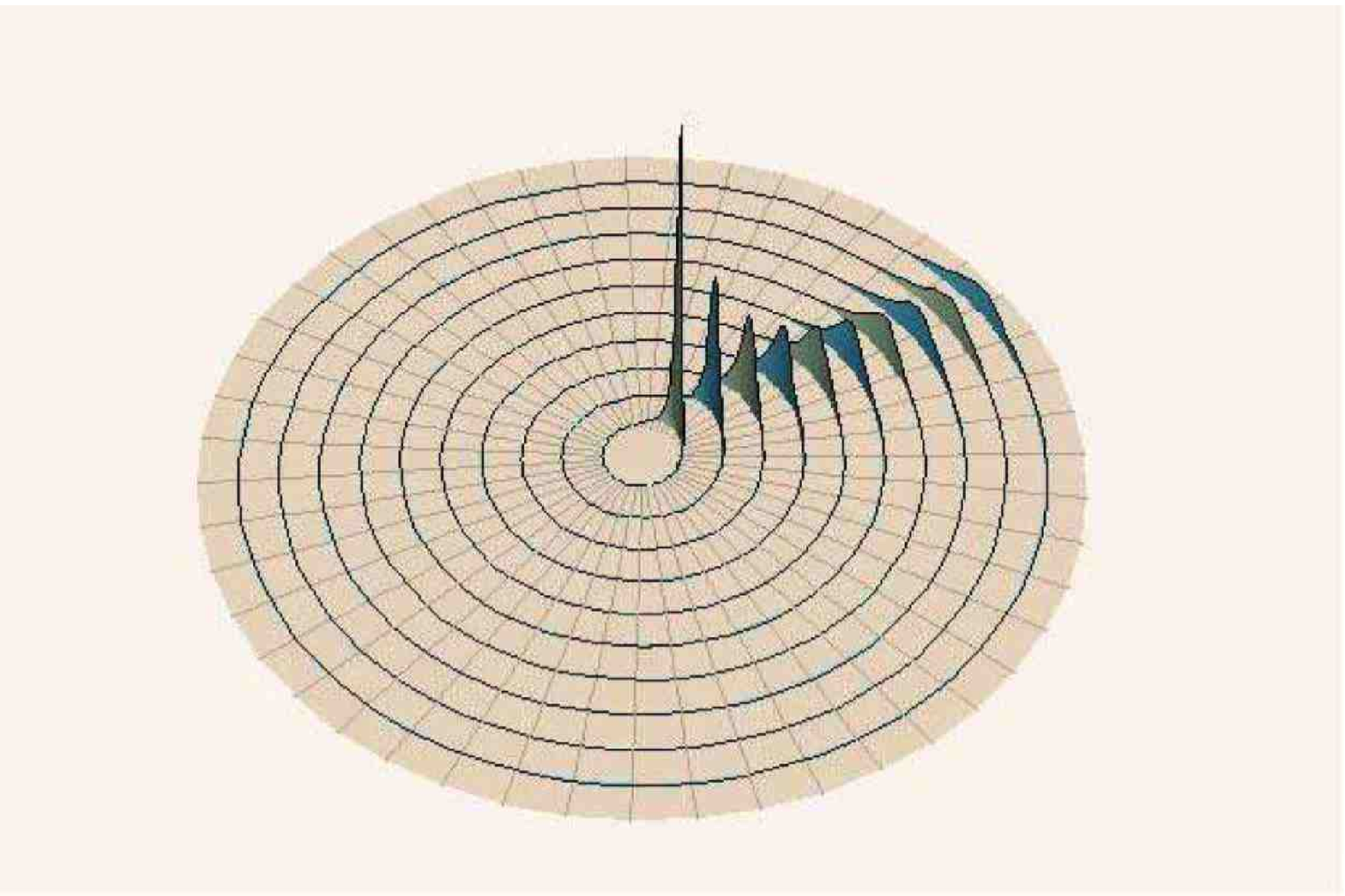}}
\caption{Time-lapse plots of $|\2G(x-iy)|$ in the far zone, showing the evolution of a \sl single pulse \rm with  propagation vector  $y=(0,0, a, b)$.  \sl Clockwise from upper left: \rm  $b/a=1.5,\ 1.1,\ 1.01,\ 1.0001$.  As $b/a\to 1$,  $y$ approaches the light cone and the pulsed beam becomes more and more focused around the ray $y$. We have taken the slice $x_2=0$, so that the disk $\5D$ becomes the interval $[-a,a]$ on the $x_1$-axis and the pulse propgates in the $x_3$ direction of the $x_1$-$x_3$ plane.}
\label{FigFar}
\end{figure}

\begin{figure}[ht]
\cl{\includegraphics[width=3.8 in]{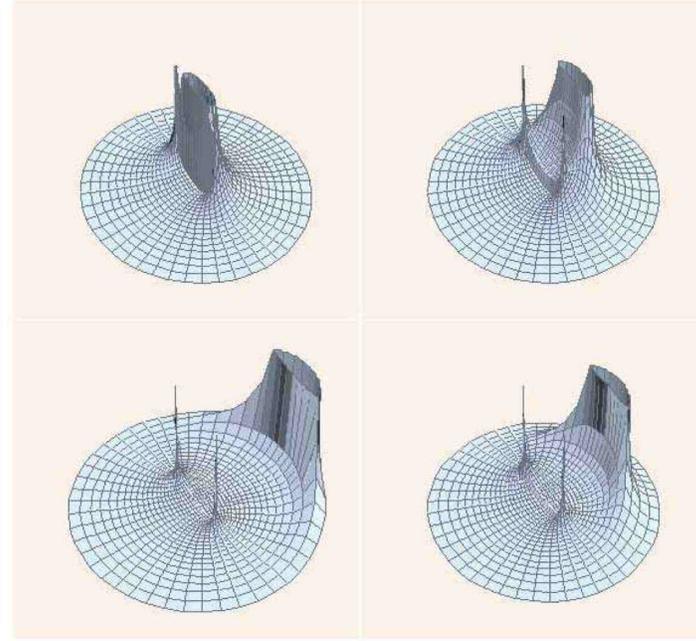}}
\caption{Near-zone graphs  of $|\2G(x-iy)|^2$ with $y=(0,0, 1, 1.01)$ immediately after launch,  evolving in the $x_1$-$x_3$ plane with $x_2=0$ as in Fig.~1. \sl Clockwise from upper left: \rm $t = 0.1,  1,  2 , 3.$ The ellipsoidal wave fronts and hyperbolic flow lines are visible. The top of the peak is cut off to show the behavior near the base. The two spikes represent the branch circle, whose slice with $x_2=0$ consists of the points $(\pm 1,0,0)$.}
\label{FigNear}
\end{figure}

The eigenwavelets have the spheroids $\5S_p$ as \sl wave fronts \rm and  propagate out along the orthogonal hyperboloids $\5H_q$ with strength decaying monotonically away from the front beam axis. Hence they have no \sl sidelobes, \rm  which makes  them potentially useful for applications to communication, radar and related areas. These properties are illustrated in Figures \ref{FigFar} and \ref{FigNear}.

We may visualize the effects of the extension $G(\3r,t)\to \2G(\bt r, \2t)$ as follows. The extension $t\to\2t$ replaces the spherical \sl impulse \rm $\d(t-r)$ in \eq{prop} by a \sl spherical pulse \rm $\2\d(\2t-r)$ of duration $|b|$. The extension  $ r\to\2r$ then \sl deforms \rm this spherical pulse to a pulsed beam in the direction of $\3a/b$. By \eq{far},
\begin{align}\lab{far2}
r\gg a\imp\2r\app r-ia\cos\q,
\end{align}
hence the larger we choose $a$, the stronger the dependence of $\2r$ on $\cos\q$ in the far zone and the more \sl focused \rm the beam.

Let us emphasize that $\2G$ depends on the complex spatial vector 
$\bt r\in\cc3$ only through the complex distance $\2r$ by writing
\begin{align}\lab{Y}
\Y(\2r,\2t)=\2G(\bt r,\2t)=\frac1{2\p i\2r(\2t-\2r)}.
\end{align}
Due to the factor $\2r$ in the denominator, $\Y$ is discontinuous across $\5D$
and singular on $\5C$. $\5D$ generalizes the point singularity of $G$ at $\3r=\30$ and will be the spatial support of the source \eq{2d}. To avoid further singularities,  the factor 
\begin{align*}
\2t-\2r=(t-p)-i(b-q)
\end{align*}
must not vanish for any $\3r$. By \eq{qa}, 
\begin{align}\lab{ba}
b-q\ne 0\ \forall \3r\iff a< |b|,
\end{align}
so a necessary and sufficient condition for $\Y(\2r, \2t)$ to be analytic whenever $\3r\notin\5D$ is that $(\bt r, \2t)\in\5T$. Recalling that the tightness of the beam is controlled by the size of $a$, \eq{ba} means that the beam cannot become tighter than a single ray and, in fact, fails to be analytic \sl along the ray \rm in the limit $a=|b|$.

The volume element in $\rr3$ in oblate spheroidal coordinates is
\begin{align}\lab{dV}
dV=\frac1a(p^2+q^2) dp\, dq\,d\f=\frac1a |\2r|^2 dp\, dq\,d\f,
\end{align}
hence $\Y$ is \sl locally \rm integrable and square integrable. A differentiation gives
\begin{align}\lab{TD}
4\p\2\d(x-iy)\=\Box_x\2G(x-iy)=0\qqq (x-iy\in\5T,\ \3r\notin\5D).
\end{align}
Therefore $\2\d(x-iy)$, with $y$ a fixed timelike vector, is a distribution in 
$x=(\3r, t)$ with \sl spatial \rm  support in $\5D$. (The \sl temporal \rm support is noncompact; in fact, $\2\d(x-iy)$ decays as $1/\2t$ due to the Cauchy kernel.)

The source $\2\d(x-iy)$ was computed explicitly in \ci{K3} and turns out to be quite singular. It consists of a single layer and a double layer on $\5D$, both of which diverge on the boundary $\5C$ where $\Y$ is singular. We will  
compute \sl regularized \rm versions of $\Y$ and $\2\d$ by using the freedom to deform the branch cut to eliminate the singularity on $\5C$.

\section{Regularization by branch cut deformation}

A  general branch cut  $\5B$ is a \sl membrane \rm  obtained by a continuous deformation of the disk $\5D$ leaving its boundary intact, \begin{align}\lab{bdy2}
\pl\5B=\5C.
\end{align}
$\5B$ inherits an orientation from $\5D$, which in turn is oriented by $\3a$. Let $V_{\5B}$ be the compact volume swept out in the deformation from $\5D$ to $\5B$. Let us define the complex distance $\2r\6{\5B}$ with branch cut $\5B$ in terms of $\2r=\2r\6{\5D}$ by
\begin{align}\lab{sigB}
\2r\6{\5B}=\begin{cases} \2r &\text{if }\3r\notin V_{\5B}\\
-\2r &\text{if } \3r\in V_{\5B}\,. \end{cases}
\end{align}
I claim that \sl $\2r\6{\5B}$ is continuous  except for a sign reversal across $\5B$, \rm  generalizing the sign reversal of $\2r$ across $\5D$. This can be seen most simply if  $\5B$ does not intersect the interior of $\5D$, so that they have only the boundary in common.
Then $V_\5B$ is either all on the `positive' or all on the `negative' side of $\5D$. If $V_\5B$ is `positive,' then its boundary is
\begin{align}\lab{Vp}
\pl V_\5B=\5B-\5D,
\end{align}
meaning that the orientation (outward normal) of the boundary is positive on $\5B$ and negative on $\5D$. Since $\2r$ changes sign upon crossing $\5D$ `upward' into $V_\5B$, \eq{sigB} shows that $\2r\6{\5B}$ is continuous across $\5D$. This proves that its only discontinuity is the sign reversal in crossing $\5B$, as claimed. Similarly, if $V_\5B$ is `negative,' then its boundary is
\begin{align}\lab{Vn}
\pl V_\5B=\5D-\5B
\end{align}
and the above argument remains valid.  To handle branch cuts that intersect the interior of $\5D$, we restate the `negative' case \eq{Vn} by declaring $V_\5B$ \sl negatively oriented, \rm so that its boundary is oriented by the \sl inward normal. \rm Denoting the negatively oriented volume by $-V_\5B$, \eq{Vn} can be restated as
\begin{align}\lab{Vn2}
\pl (-V_\5B)=\5B-\5D.
\end{align}
Hence the rule \eq{Vp} applies to every branch cut $\5B$ obtained by a continuous deformation of $\5D$, whether or not it intersects the interior of $\5D$, provided the orientation of the swept-out volume is taken into account. If $\5B$ intersects the interior of $\5D$, then $V_\5B$ has both positively and negatively oriented components. The definition \eq{sigB} of the branch $\2r\6{\5B}$ remains valid whether $V_\5B$ (or any of its components) is positively or negatively oriented. 

\begin{figure}[ht]
\begin{center}
\includegraphics[width=3 in]{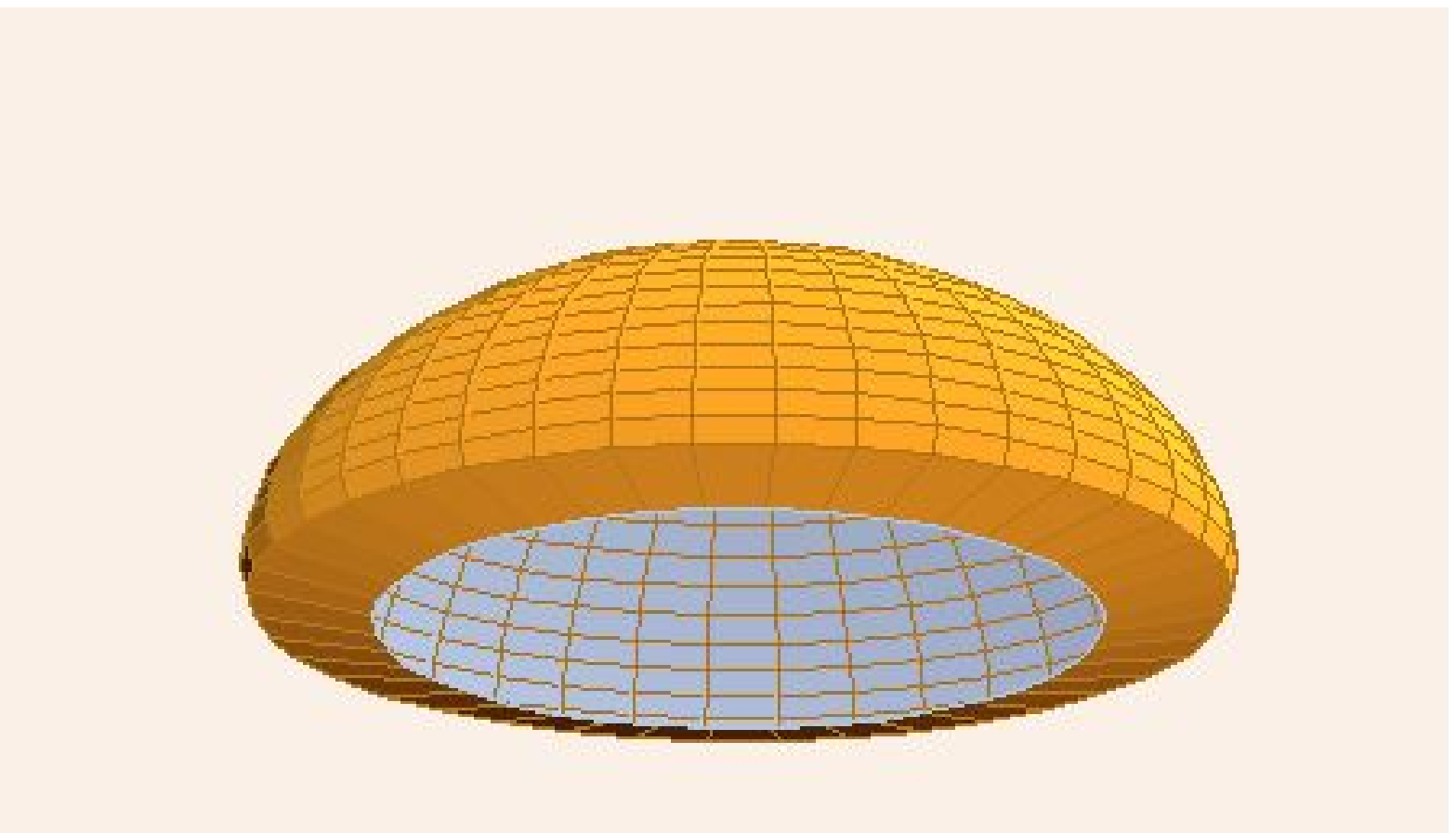}
\caption{The upper hemispheroidal branch cut $\5B\9+_\a$ with its apron.}
\label{FigOScu}
\end{center}
\end{figure}

Of special interest will be the \sl upper and lower spheroidal branch cuts \rm
\begin{align}\lab{Bp}
\5B\9\pm_\a=\5S\9\pm_\a\cup\5A_\a
\end{align}
where $\5S\9\pm_\a$ denote the upper and lower hemispheroids
\begin{align*}
\5S\9\pm_\a=\{\3r\in\5S_\a: \pm z>0\}
\end{align*}
and
\begin{align*}
\5A_\a=\{\3r:\ \3r\cdot \3a=0,\ a^2\le r^2\le \a^2+a^2\}
\end{align*}
is the \sl apron \rm connecting them to $\5C$, which must be included so that
$\pl\5B\9\pm_\a=\5C$ as required.
The cut $\5B\9+_\a$ is depicted in Figure \ref{FigOScu}.

We can now construct a \sl regularized \rm version of the extended propagator
$\Y$ by taking the \sl average \rm of the propagators with cuts $\5B\9+_\a$ and $\5B\9-_\a$.  Denote the complex distances with cuts $\5B\9\pm_\a$ by $\2r\6\pm$ instead of $\2r\6{\5B\9\pm_\a}$, and let 
\begin{align}\lab{YA}
\Y_A(\2r, \2t)=\frac12\LB\Y(\2r\6+,\2t)+\Y(\2r\6-,\2t)\RB=\2G_A(x-iy).
\end{align}
Let $V_\a\9\pm$ be the interiors of the upper and lower hemispheroids. By 
\eq{sigB}, 
\begin{align}
\3r\in V_\a\9+&\imp\2r\6+=-\2r,\ \  \2r\6-=\2r\\
\3r\in V_\a\9-&\imp\2r\6+=\2r,\  \ \2r\6-=-\2r.
\end{align}
Hence, in both $V_\a\9\pm$ we have
\begin{align}\lab{Yint}
\Y_A(\2r, \2t)=
\frac1{4\p i\2r(\2t-\2r)}-\frac1{4\p i\2r(\2t+\2r)}=\frac1{2\p i(\2t^2-\2r^2)}\,,
\end{align}
which is \sl independent \rm of the choice of branch. This shows that the discontinuities across the aprons cancel in the average $\Y_A$. Furthermore, by \eq{ba} we have
\begin{align*}
|b|>a\imp\2t^2-\2r^2=(\2t-\2r)(\2t+\2r)\ne 0,
\end{align*}
showing that the singularities on $\5C$ cancel as well. That is, \sl $\Y_A$ is analytic at all interior points of the spheroid $\5S_\a$. \rm

In the \sl exterior \rm of $\5S_\a$ we have $\2r\6\pm=\2r$ and hence $\Y_A=\Y$. Since $\5D$ is contained in $\5S_\a$ and $\Y$ is analytic outside of $\5D$, we conclude that $\Y_A(\2r, \2t)$ fails to be analytic only when $\3r\in\5S_\a$. Denoting the  interior field by $\Y_1$ and the exterior  field by $\Y_2$, we have
\begin{align}\lab{Ypm}
\Y_1(\2r,\2t)&=\frac12\LB\Y(\2r,\2t)+\Y(-\2r,\2t)\RB\\
\Y_2(\2r,\2t)&=\Y(\2r,\2t).\nt
\end{align}
Thus $\Y_A$ is analytic except for a \sl bounded jump discontinuity \rm across $\5S_\a$ given by
\begin{align}\lab{dYa}
\Y_J(\2r,\2t)\=\Y_2(\2r,\2t)-\Y_1(\2r,\2t)
=\frac12\LB\Y(\2r,\2t)-\Y(-\2r,\2t)\RB=\2G_J(x-iy).
\end{align}
It follows  by the same argument as in \eq{TD} that the source distribution 
\begin{align}\lab{2da}
4\p\2\d_A(x-iy)\=\Box_x\Y_A(\2r, \2t)
\end{align}
is supported spatially on $\5S_\a$. Because $\2\d_A$ is obtained by twice differentiating a discontinuous function, it consists of a combination of single and double layers on $\5S_\a$. But the jump discontinuity in $\Y_A$ is \sl bounded \rm (unlike that in $\Y$, which diverges on $\5C$), and so are these layers; see \ci{K4}.

The above arguments remain valid if instead of $\5B\9\pm_\a$ we use \sl any \rm two branch cuts whose common interior $V$ contains the branch circle $\5C$. In that case, the averaged propagator is analytic in $\5T$ except for a finite discontinuity when $\3r$ crosses the boundary $\pl V$, and its source distribution is supported spatially on $\pl V$. However, the above choice has the advantage that  $\pl V=\5S_\a$ are \sl wave fronts, \rm  hence all parts of the surface radiate simultaneously and coherently.

\section{Extended Huygens sources}

Let $ H$ be the Heaviside step function. Since $0\le p<\a$ in the interior of $\5S_\a$ and $p>\a$ in the exterior, we have
\begin{align}\lab{Ya}
\Y_A(\2r, \2t)= H(\a-p)\Y_1(\2r, \2t)+ H(p-\a)\Y_2(\2r, \2t)
\end{align}
where the interior and exterior fields  are given by \eq{Ypm}. This can be used to compute the source distribution $\2\d_A$ defined in \eq{2da}, and the result is sum of terms with factors $\d(p-\a)$ and $\d'(p-\a)$. The former are interpreted as \sl single layers \rm on $\5S_\a$, and the latter as \sl double layers. \rm 

An interesting practical question is whether the wavelets $\Y_A$, interpreted as acoustic pulsed beams, can be \sl realized \rm by manufacturing their sources. A similar question can be posed for their electromagnetic counterparts, which solve Maxwell's equations; see \ci{K4}. It is doubtful whether an acoustic source can be produced including double layers, and the problem becomes even more difficult in the electromagnetic case because the current density involves yet another derivative, hence a still higher layer \ci{K4a}. The multilayered structure is unavoidable as long as we insist on \sl surface sources. \rm We now propose a method for constructing solutions of the wave equation  where the transition occurs in a \sl shell \rm instead of a surface. It will be simpler to present this method initially in a somewhat more general context.

Given a function $p(\3r,t)$ on $\4R^{n,1}$ and two regular values $p_1<p_2$ in its range, define two time-dependent surfaces and volumes in $\rr n$ by
\begin{align*}
S_1\0t&=\{\3r: p(\3r, t)=p_1\},\  S_2\0t=\{\3r: p(\3r,t)=p_2\} \\
V_1\0t&=\{\3r: p(\3r, t)<p_1\},\  V_2\0t=\{\3r: p(\3r,t)>p_2\}.
\end{align*}
Let $f_1, f_2$ be solutions of the wave equation in $\4R^{n,1}$
with sources $g_1, g_2$:
\begin{align}\lab{fk}
\Box f_k(\3r,t)=g_k, \qq k=1,2. 
\end{align} 
We want to construct an \sl interpolated solution \rm $f(\3r,t)$ such that
\begin{align}\lab{inter}
f(\3r,t)=f_k(\3r,t)\ \forall\3r\in V_k\0t
\end{align}
and compute its source. This can be done by choosing functions $h_k(\3r,t)$ with
\begin{align}\lab{hk}
h_1(\3r,t)=\begin{cases} 1, &\3r\in V_1\0t\\0, & \3r\in V_2\0t\end{cases},
\qqq h_2(\3r,t)=1-h_1(\3r,t)
\end{align}
and letting
\begin{align}\lab{f}
f=h_1 f_1+h_2 f_2\=h_k f_k
\end{align}
where the (Einstein) summation convention is used. The source of $f$ is found to consists of two parts,
\begin{align}\lab{g}
g=\Box f=g\6I+g\6T\,,
\end{align}
where 
\begin{align}\lab{gI}
g\6I=h_k g_k
\end{align}
is an \sl interpolated source \rm and 
\begin{align}\lab{gT}
g\6T=2\dot h_k\dot f_k-2\grad h_k\cdot\grad f_k+(\Box h_k) f_k
\qqq(\dot f\=\pl_t f)
\end{align}
is a \sl transitional source \rm which,  by \eq{hk}, is supported on the  transition shell 
\begin{align}\lab{VT}
V_T\0t=\{\3r: p_1\le p(\3r,t)\le p_2\}
\end{align}
and depends only on the \sl jump field \rm $f\6J=f_2-f_1$:
\begin{align}\lab{gT2}
g\6T=2\dot h_2\dot f\6J-2\grad h_2\cdot\grad f\6J+(\Box h_2) f\6J.
\end{align}
Now suppose that $V_1\0t$ and $V_T\0t$ are compact and we are given only one
source $g_2$, supported in $V_1\0t$. Letting $f_2$ be its causal field, our objective is to find an \sl equivalent source \rm $g$ supported in $V_T\0t$ whose causal field $f$ is identical with $f_2$ in $V_2\0t$. It suffices to choose \sl any \rm solution $f_1$ whose source $g_1$ is supported in $V_2\0t$, since the interpolated source \eq{gI} then vanishes and hence $g=g\6T$.  $f_1$ is a sourceless \sl internal field \rm in $V_1\0t$, and  the source $g\6T$ so constructed on $V_T\0t$ generalizes the idea of a \sl Huygens source \rm on a surface surrounding the support of $g_2$.
We may recover the latter by assuming that $p$ is time-independent (hence so are $S_k$ and $V_k$) and choosing $h_k(\3r)$ so that
\begin{align*}
\lim_{p_1\to p_2}\grad h_2(\3r)=\d(p(\3r)-p_2)\3n(\3r)
\end{align*}
where $\3n(\3r)$ is a field of orthogonal vectors on $S_2$ pointing into $V_2$. The corresponding scheme in the electromagnetic case reduces to the usual boundary conditions on an interface between two media \ci{K4a}.

Returning to $n=3$ with $p=\re\2r$, let $f_k=\Y_k$ as in \eq{Ypm} and $h_k$ be time-independent (\eg  functions of $p$ only). A \sl smoothed \rm version of $\Y_A$ \eq{YA} is
\begin{align}\lab{2YA}
\Y_A^{\,\rm sm}=h_1\Y_1+h_2\Y_2.
\end{align}
Since $\Y_k$ are sourceless in $V_T$, \eq{g} gives the smoothed version of  \eq{2da} as
\begin{align}\lab{2SA}
4\p\2\d_A^{\,\rm sm}=\Box_x\Y_A^{\,\rm sm}
=-2\grad h_2\cdot\grad \Y_J-(\D h_2) \Y_J
\end{align}
where
\begin{align*}
\Y_J=\Y_2-\Y_1=\frac12\LB\Y(\2r,\2t)-\Y(-\2r, \2t)\RB
=\frac{\2t}{2\p i\2r(\2t^2-\2r^2)}
\end{align*}
is the jump field from $V_1$ to $V_2$ as in \eq{dYa}, but no longer restricted to a single spheroid $\5S_\a$.  If we now let $p_1\to p_2=\a$ and
\begin{align*}
h_1=H(\a-p),\qq h_2=H(p-\a),
\end{align*}
then the transition becomes abrupt on $\5S_\a$ and $\Y_A^{\,\rm sm}$ becomes $\Y_A$ \eq{YA}. Since
\begin{align*}
\grad h_2&=\d(p-\a)\grad p\\
\D h_2&=\d'(p-\a)| \grad p |^2+\d(p-\a)\D p,
\end{align*}
equation \eq{2SA} becomes
\begin{align*}
4\p\2\d_A=-2\d(p-\a)\grad p\cdot\grad\Y_J-\d(p-\a)\D p \,\Y_J
-\d'(p-\a)| \grad p |^2\Y_J
\end{align*}
displaying the aforementioned single and double layer structure on $\5S_\a$. 
To get an explicit expression, use  \ci[Appendix]{K4} 
\begin{align*}
&\grad p= \frac{p\3r+q\3a}{p^2+q^2}, \qqq\  \D p=\frac{2p}{p^2+q^2}\\
& |\grad p|^2=\frac{p^2+a^2}{p^2+q^2},\qqq 
\grad p\cdot\grad q=0
\end{align*}
and
\begin{align*}
\grad p\cdot\grad\Y_J&=\Y_J'\grad p\cdot\grad\2r=\Y_J' | \grad p|^2
\end{align*}
where $\Y_J'$ is the complex derivative of $\Y(\2r,\2t)$ with respect to $\2r$ (keeping in mind that $\Y(\pm\2r, \2t)$ are analytic in $\2r$ for $p>0$),
\begin{align*}
\Y_J'=\frac{\pl\Y_J}{\pl\2r}=-\frac{\2t}{2\p i\2r^2(\2t^2-\2r^2)^2}.
\end{align*}

\section{Conclusions}
Although I have concentrated on the wave equation in four-dimensional Minkowski space $\4R^{3,1}$, similar considerations apply in $\4R^{n,1}$. In fact, the awkward extension of the propagator, using the Cauchy kernel in time but the complex distance in space, becomes much more natural when 
$\2G(\bt r, \2t)$ is viewed as the retarded part of the analytic continuation of the fundamental solution  $G_E(\3R)$ of Laplace's equation in \sl Euclidean \rm $\4R^{n+1}$ \ci{K0,K3}, based on the complex distance 
\begin{align*}
\2R=\sr{\bt R\cdot\bt R}, \qqq \bt R\in\4C^{n+1},
\end{align*}
whose branch points form a sphere $S^{n-1}$ in $\4R^{n+1}$ of codimension 2 and radius $|\im\bt R|$. The extended delta function $\2\d\6E(\bt R)$,\footnote{The subscript distinguishes $\2\d\6E(\bt R)$ from the \sl Minkowskian \rm $\2\d(\2x)$ in \eq{2d}.}  defined by applying the Laplacian in $\3R$ to the extension $\2G_E(\bt R)$, is supported on $S^{n-1}$ for \sl odd \rm $n\ge 3$, but a branch cut, consisting of a `membrane' bounded by $S^{n-1}$, is needed in all other cases.\footnote{This is because $G_E(\3R)=c_n/R^{n-1}$ for $n\ge 2$ and $G_E=c_1\log R$ for $n=1$.}
Given any test function $f$ in $\4R^{n+1}$, the convolution
\begin{align}\lab{f3}
\2f(\bt R)=\int_{\4R^{n+1}}\2\d\6E(\bt R-\3R') f(\3R')\, dV(\3R')
\end{align}
defines an extension of $f$ to $\4C^{n+1}$, non-holomorphic in general, whose restriction to the \sl Minkowski \rm subspace $\4R^{n,1}$, obtained by letting $\bt R=(\3r, it)$, is a solution of the following initial-value problem for the wave equation:
\begin{align}
&(\pl_t^2-\D_{\3r})\2f(\3r, it)=0\lab{hom2}\\
&\2f(\3r, 0)=f(\3r, 0)\lab{init1}\\
&(\pl_t-i\pl_b)\2f(\3r, b+ it)\mid_{b=t=0}=0.\lab{init2}
\end{align}
For odd $n\ge 3$, the proof of \eq{hom2} is based on the fact that $\2\d\6E$ is distributed uniformly on  $S^{n-1}$ and hence $\2f$ is a \sl spherical mean \rm of $f$ \ci{J55}. This relates the support of $\2\d\6E$ for odd $n\ge 3$ to \sl Huygens principle. \rm The other cases can be treated by applying a distributional version of Hadamard's method of descent.

Equation \eq{init2} states that $\2f(\3r, b+it)$ satisfies the Cauchy-Riemann equation in its last variable; but since this holds only at one point, it does not imply analyticity ---  as it cannot since $f(\3r, b)$ need not have any analytic continuation in $b$. If one exists, it is indeed given by $\2f(\3r, b+it)$. This generalizes an old theorem by Paul Garabedian \ci[pp 191--202]{G64}. 

\section*{Acknowledgements}
I thank Dr.~Arje Nachman for his sustained support of my research, most recently through AFOSR Grant  
\#FA9550-04-1-0139.

\end{document}